# Magnetoelastic coupling in the stretched diamond lattice of TbTaO$_4$

Xiaotian Zhang,[a] Nicola D. Kelly,[ab] Denis Sheptyakov,[c] Cheng Liu,[a] Shiyu Deng,[ad]

Siddharth S. Saxena[a] and Siân E. Dutton *[a]

The magnetic structure of diamond-like lattice has been studied extensviely in terms of the magnetic frustration. Here we report the distortion of stretched diamond lattice of Tb$^{3+}$ (4$f^8$) in $M$-TbTaO$_4$ on application of a magnetic field. We have investigated the structural and magnetic properties of $M$ phase terbium tantalate $M$-TbTaO$_4$ as a function of temperature and magnetic field using magnetometry and powder neutron diffraction. Sharp λ-shape transitions in $d(\chi T)/dT, dM/dH$ and specific heat data confirm the previously reported three-dimensional (3D) antiferromagnetic ordering at $T_N$~2.25 K. On application of a magnetic field the Néel temperature is found to decrease and variable field neutron diffraction experiments below $T_N$ at 1.6 K show an increase in both the bond and angle distortion of the stretched diamond lattice with magnetic field, indicating a potential quantum magneto-elastic coupling effect. By combining our magnetometry, heat capacity and neutron diffraction results we generate a magnetic phase diagram for $M$-TbTaO$_4$ as a function of temperature and field.

## Introduction

The previous studies of magnetism in diamond-like lattices has been mainly focused on the ceramic materials, including the magnetic A-site spinels, NiRh$_2$O$_4$[1], CuRh$_2$O$_4$[2] and MAl$_2$O$_4$[3] (M=Co, Fe, Mn) as well as scheelite KRuO$_4$[4] and lanthanide NaCeO$_2$[5] materials. The undistorted diamond lattice, a repeating pattern of centred tetrahedra, is one of the bipartite lattices capable of exhibiting colinear antiferromagnetic interactions between the nearest-neighbour spins ($J_1$) and the quantum ground state has magnetic long range Néel order ("up-down")[6]. However, most reported materials tend to show strong magnetic frustration, since the nearest neighbour interaction $J_1$ is relatively weak compared to the next-nearest neighbour interaction $J_2$[6]. This gives rise to a variety of magnetic phenomena, ranging from long-range ordered states to disordered spin liquid and spin glass states[7, 8] and topologic paramagnetism[9, 10]. The competition between $J_1$ and $J_2$ can also result in multiple low energy magnetic regimes and a complicated phase diagram[11].

When the diamond lattice is distorted by symmetry lowering the related stretched diamond lattice is generated. In 2021, Bordelon et al. used a "stretched" (distorted) diamond lattice framework to explain the $J_1$-$J_2$ interaction in the tetragonal spinel LiYbO$_2$. The magnetic order of the Yb$^{3+}$ ions becomes commensurate on application of a magnetic field[12]. However, the reported spin spiral magnetic structure in zero field is still subject to debate[13-15]. In 2022, Kelly et al. reported the magnetic lanthanide ions ($Ln^{3+}$) in monoclinic ferguson-type $Ln$TaO$_4$ also form a stretched diamond lattice and introduced the concepts of bond and angular distortion to quantify the distortions in the stretch diamond lattice[16].

The rare-earth tantalates $Ln$TaO$_4$ [$Ln$= Y, La–Lu] have attracted increasing attention due to their wide application, such as phosphors[17], thermal barrier[18], scintillators[19] and dielectric ceramics[20]. They adopt a number of different structural polymorphs depending on the synthetic conditions[20-25]. The magnetic $Ln^{3+}$ ions form a stretched diamond network in both the low temperature M ($I2/a$, monoclinic, fergusonite) and high temperature T ($I4_1/a$, tetragonal, scheelite) phases[16]. Prior work on $M$-$Ln$TaO$_4$ powders, mainly focused on the luminescent and thermal properties[18, 19, 26, 27] rather than magnetism. More magnetic studies have been done on the isostructural niobates, $Ln$NbO$_4$[28, 29], potentially due to their lower synthesis temperature. In 1996, Tsunekawa et al. reported the magnetic susceptibilities of NdTaO$_4$, HoTaO$_4$ and ErTaO$_4$ single crystals with negative Curie-Weiss temperatures and no magnetic transitions between 4.2 and 300 K[30]. Recently, Kelly et al. reported the bulk magnetisation of polycrystalline $M$-$Ln$TaO$_4$ ($Ln$= Nd, Sm-Er, Y) samples. In agreement with previous work all were found to have negative Curie-Weiss temperatures and no compounds order above 2 K, except $M$-TbTaO$_4$ with an antiferromagnetic transition at 2.25 K. Powder neutron

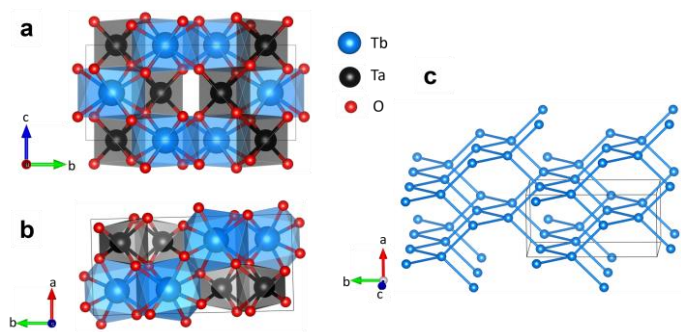

Fig. 1 (a-b) Monoclinic crystal structure of $M$-TbTaO$_4$ (space group $I2/a$). (c) the distorted diamond lattice of Tb site. The unit cell is shown in thin grey lines.

diffraction (PND) was used to determine its magnetic structure, revealing that it forms a commensurate AFM structure with $\vec{k}=0$[16].

Here, we expand on the prior work of Kelly et al. and focus on the nuclear and magnetic structure of $M$-TbTaO$_4$ at variable temperature and magnetic field using powder neutron diffraction. At 1.6 K, below $T_N$, a slight increase in angle distortion and band distortion is observed from 0 to 6 T. We interpret this as antiferromagnetic ordering triggering a quantum magneto-elastic coupling in $M$-TbTaO$_4$ due to the presence of $4f^8$ Tb$^{3+}$ which are known to exhibit quantum behaviour[31]. Our variable temperature and field magnetic susceptibility, and specific heat measurements allow us to track changes in the magnetism. From these measurements, the Néel temperature of $M$-TbTaO$_4$ is found to be suppressed by the magnetic field and a transition to a canted antiferromagnetic state is observed on application of a magnetic field.

## Experimental

Polycrystalline $M$-TbTaO$_4$ was synthesised according to a solid-state ceramic reaction as has been reported elsewhere[32]. Ta$_2$O$_5$ (Alfa Aesar, 99.993%) and Tb$_4$O$_7$ (Alfa Aesar, 99.9%) powders were firstly heated at 800 °C overnight to remove moisture. Then 1:1 molar amounts of the reagents were thoroughly mixed in an agate pestle and mortar, pressed into a 7-mm pellet and placed in an alumina crucible. The pellets were heated for 72 h at 1500 °C in air with intermediate regrinding every 24 h.

Powder x-ray diffraction (PXRD) was carried out at room temperature on a Bruker D8 diffractometer (Cu K$\alpha$, $\lambda$=1.541 Å) in the range $10 \leq 2\theta(°) \leq 70$ with a step size of 0.02 °, 0.6 s per step. Rietveld refinements[33] were carried out using TOPAS[34] with a Chebyshev polynomial background and Thompson-Cox-Hastings pseudo-Voigt peak shape[35]. VESTA[36] was used for crystal structure visualization and production of figures.

Powder neutron diffraction (PND) was carried out on an 8 g sample of $M$-TbTaO$_4$, prepared by combining two batches confirmed to be phase pure by PXRD. The sample was pressed into disc-shape with a diameter of 7.1 mm and enclosed within the cadmium container. Cadmium platelets were also placed between discs, ensuring they remained immobilized. PND was conducted on the High-Resolution Powder Diffractometer for Thermal Neutrons (HRPT), Paul Scherrer Institut (PSI), Villigen, Switzerland [37], using an Orange cryostat (1.5 $\leq T$ (K) $\leq$ 300). Neutrons with $\lambda$= 2.4487(2) Å were obtained by using the (400) reflection on the focusing Ge monochromator at a take-off angle of 120 deg. The determination of the magnetic structure was carried out using TOPAS[34]. The background was modelled with a Chebyshev polynomial, and the peak shape was modelled with a modified Thompson-Cox-Hastings pseudo-Voigt[35] function with axial divergence asymmetry.

The dc magnetisation was measured on warming on a Quantum Design MPMS®3 in the temperature range 1.8 $\leq$ T (K) $\leq$ 300 under different magnetic fields ranging from 500 Oe to 70000 Oe, after cooling from 300 K in zero field (ZFC). The isothermal magnetisation was measured on the same system in the field range $\mu_0$H = 0 -7 T at different temperatures.

Heat capacity of $M$-TbTaO$_4$ was measured on a Quantum Design PPMS® DynaCool in the range 1.8 $\leq T$ (K) $\leq$ 30 under different magnetic fields ranging from 0 Oe to 70000 Oe. The sample was mixed with an equal mass of Ag powder to improve thermal conductivity and pressed into a 5 mm pellet before mounting on the sample stage with Apiezon N grease. Fitting of the relaxation curves was done using the two-tau model. The contribution of Ag to the total heat capacity was subtracted using scaled values from the literature[38]. The TbTaO$_4$ lattice contribution was estimated and subtracted using a Debye model with $\theta_D$ = 370 K[39], as is shown in Fig. S1.

## Results

### Structure

The structure of $M$-TbTaO$_4$ has been previously reported to be monoclinic fergusonite, space group 15, $I2/a$ (unconventional unit cell) or $C2/c$ (standard unit cell)[16, 22]. The Ta$^{5+}$ ions are connected by four shorter and two longer Ta-O bonds, forming an octahedron with second order Jahn-Teller distortion. The two edges are shared by neighbouring Ta-O octahedra, and

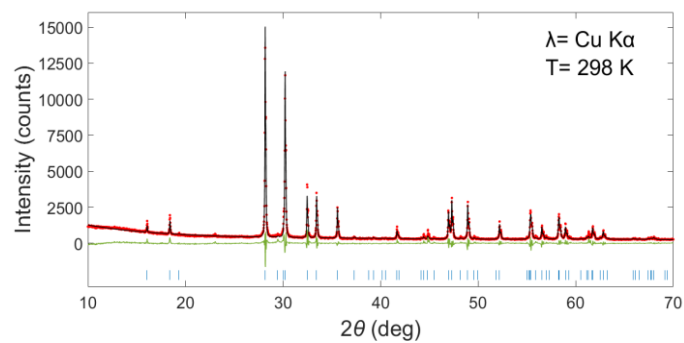

Fig. 2 Room-temperature PXRD pattern for $M$-TbTaO$_4$: red dots, experimental data; black line, calculated intensities; green line, difference pattern; blue tick marks, Bragg reflection positions.

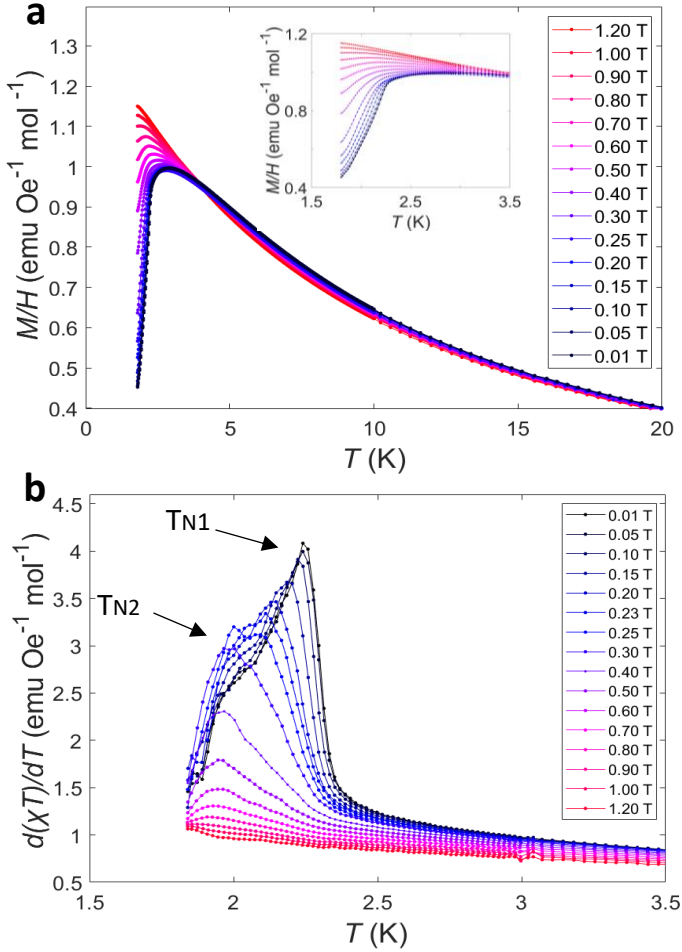

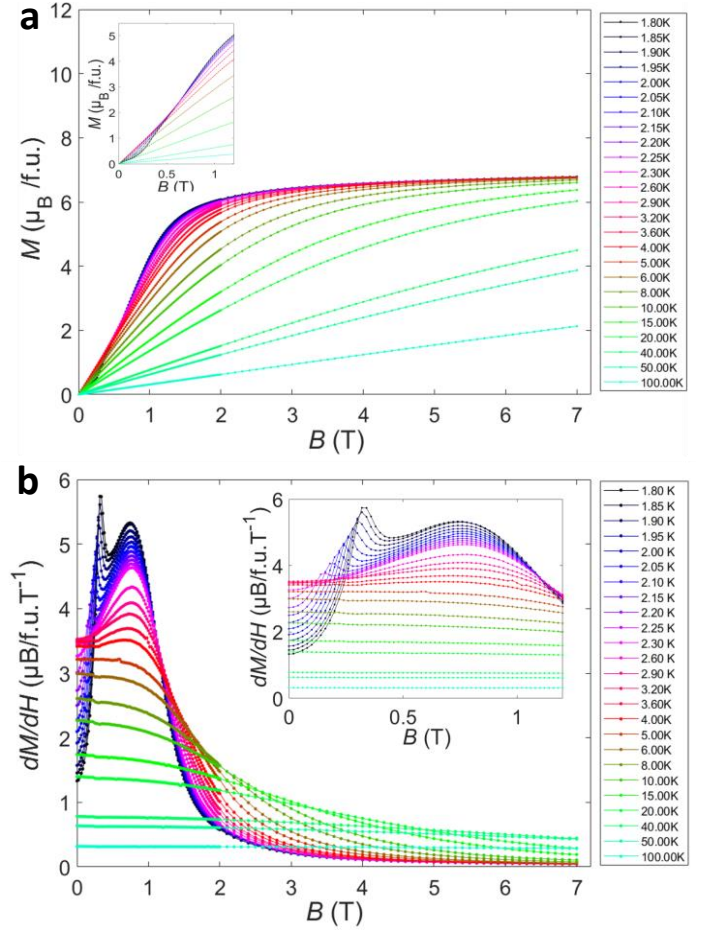

Fig. 3 (a) The ZFC magnetic susceptibility (Inset: at low temperatures) and (b) $d(\chi T)/dT$ ($\chi \approx M/H$) for $M$-TbTaO$_4$ as a function of temperature in selected fields.

Fig. 4 (a) Isothermal magnetization of $M$-TbTaO$_4$ at selected temperatures from 0-7 T (Inset: from 0-1.2 T) and (b) The corresponding derivative, $dM/dH$ as a function of field.

other edges are shared with Tb-O dodecahedra as shown in Fig. 1a and 1b. The Tb$^{3+}$ ions in the ferguosonite crystal form a stretched diamond lattice or distorted honeycomb-like structure (Fig. 1c) which is predicted to host the exotic magnetic ground state[16].

Rietveld refinement of the room-temperature PXRD of $M$-TbTaO$_4$, Fig. 2, indicated the formation of single monoclinic $M$ phase in our sample after 72 hours of sintering at 1500 °C. The unit cell parameters and the Tb$^{3+}$ and Ta$^{5+}$ atomic positions were refined, while the atomic positions of the O$^{2-}$ ions were fixed based on the neutron diffraction data from previous experiment at ILL[16, 40] (Table S1). The refined unit cell parameters and atomic positions, with no Tb$^{3+}$/Ta$^{5+}$ site disorder observed, are consistent with previous literature[18, 22].

**Magnetic susceptibility**

The ZFC magnetic susceptibility, $\chi \approx M/H$, as a function of temperature is shown in Fig. 3a under selected magnetic fields. At 0.01 T, a sharp cusp feature was observed at ~2.9 K, which agrees with the susceptibility data reported by Kelly et al[16]. The cusp temperature decreases from 2.91(8) to 2.08(8) K upon increasing the field from 0.01 to 0.80 T. Above 0.80 T, the AFM ordering is gradually evolving to a FM-like ordering where $\chi$ saturates at low temperature. At high temperatures, the magnetic susceptibility is independent of field and fits the modified Curie-Weiss law:

$$\chi - \chi_0 = \frac{C}{T - \theta_{Cw}}$$

Fitting to the Curie-Weiss law was carried out using the data collected at 1 T for T > 50 K. The effective magnetic moment was calculated from the experimental data using $\mu_{eff}/\mu_B = \sqrt{8C}$ and compared with the theoretical moment $g_J\sqrt{J(J+1)}$. The effective moment inferred from the fit is 10.23 (1) $\mu_B$/Tb is of the same order of the theoretical value 9.72 $\mu_B$/Tb, and the Weiss temperature is -9.67 (7) K consistent with that reported by Kelly et al[16].

The derivative of the product $\chi T$ with respect to temperature, $d(\chi T)/dT$ as a function of temperature at selected fields are in

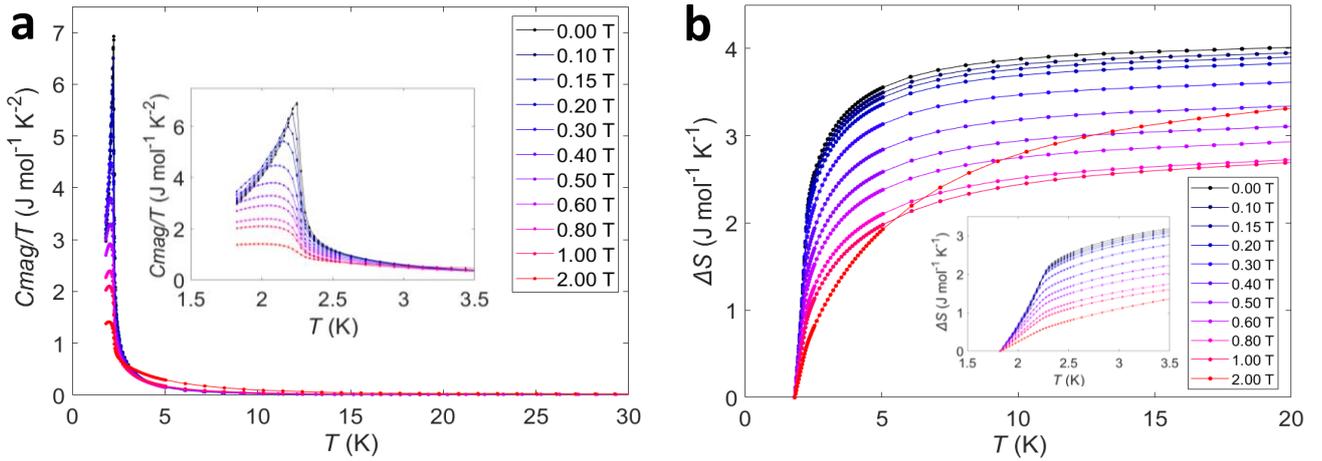

Fig. 5 (a) Magnetic heat capacity of *M*-TbTaO$_4$ as a function of temperature at selected fields and (b) magnetic entropy obtained by integration of $C_{mag}/T$. Inset: data obtained at low temperatures.

shown in Fig. 3b. At 0.01 T, a sharp lambda-type feature is observed at 2.25(1) K, which corresponds to the Néel temperature ($T_{N1}$), this indicates the onset of three-dimensional AFM ordering. Additionally, another shoulder like feature is visible at ~2.02(2) K, denote as $T_{N2}$. This secondary transition suggests a more complex magnetic structure, such as multiple AFM sublattices or competing interactions. $T_{N1}$ decreases from 2.25 (1) to 2.05(9) K upon increasing field from 0.01 to 0.30 T and $T_{N2}$ decreases from ~2.02(2) to 1.92(6) K as field increases from 0.01 to 0.80 T. This suggest that the AFM ordering is supressed on application of a magnetic field.

**Isothermal magnetisation**

Isothermal magnetisation measurements at selected temperatures are shown in Fig. 4a. Below the higher Néel temperature ($T_{N1}$), the isothermal magnetisation exhibits a characteristic 'S' shape curvature (Fig. 4a inset), indicative of complex interactions and spin re-orientation within the antiferromagnetic structure. As the field increases, the magnetisation reaches saturation by approximately 1.5 T. The resulting saturation magnetisation ($M_{sat} \approx 6~\mu_B$ /f.u.) is lower than the 9 $\mu_B$ /f.u. expected for a Heisenberg-like Tb$^{3+}$ system, reflecting spin anisotropy and powder-averaging effects. In Tb$^{3+}$

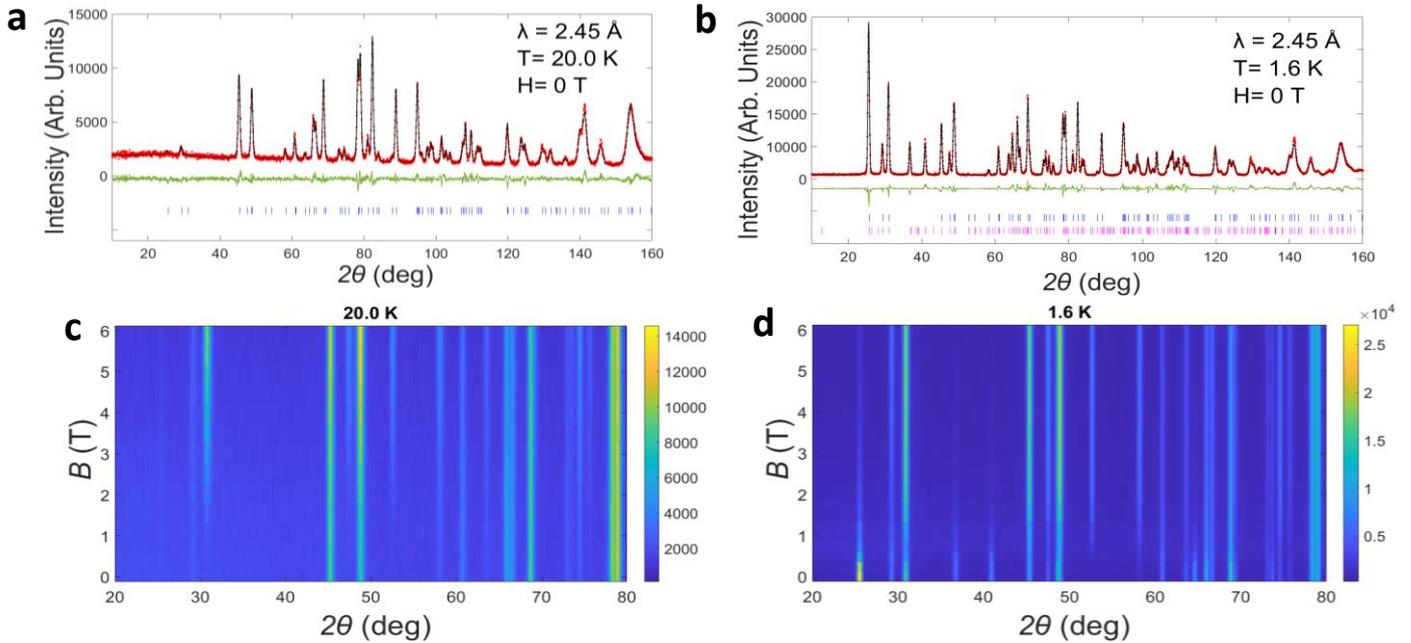

Fig. 6 (a,b) Refined PND data of *M*-TbTaO$_4$ collected at 20.0 K and 1.6 K respectively with 0 T; (c,d) The corresponding heat map showing the evolution of magnetic phase with increasing magnetic field from 0 to 6 T. Red dots, experimental data; black line, calculated intensities; green line, difference pattern; tick marks, nuclear (blue), magnetic (pink) Bragg reflection positions.

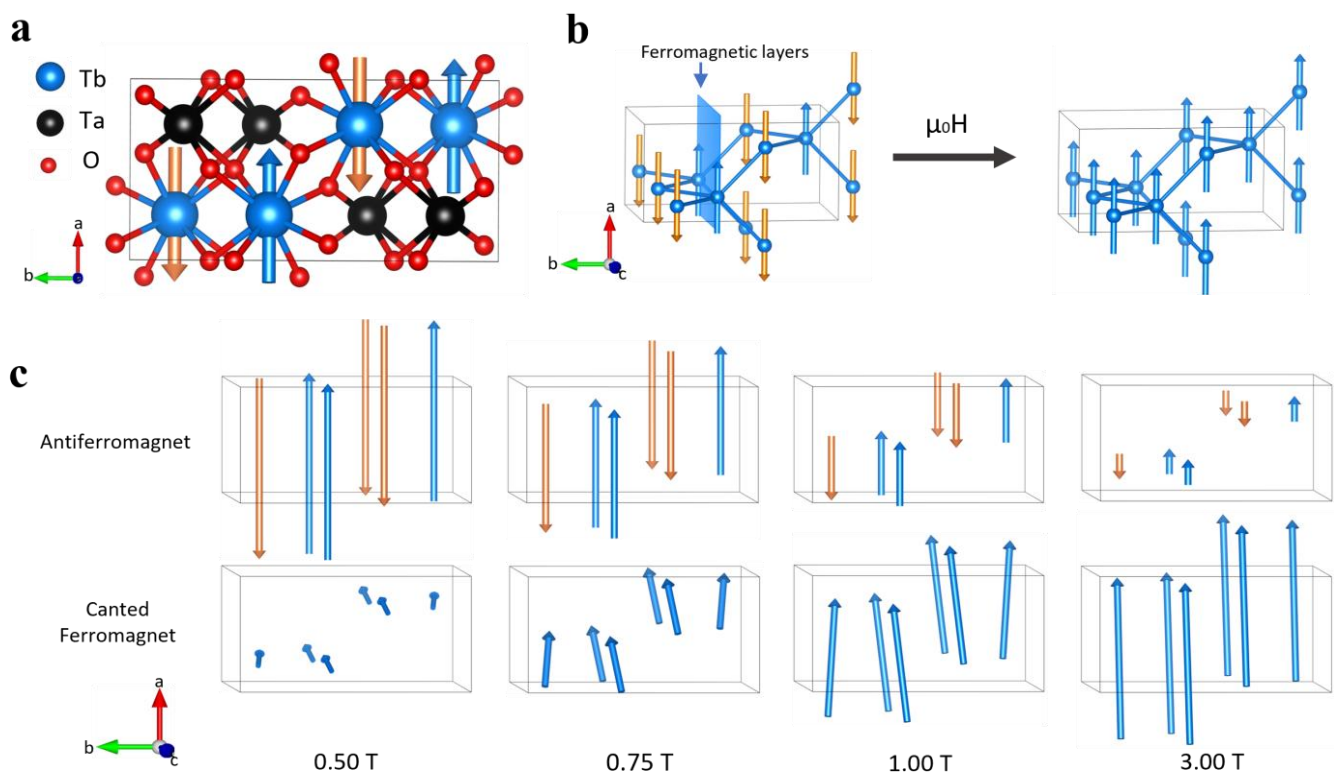

Fig. 7 (a) Magnetic structure of *M*-TbTaO$_4$ from Rietveld refinement at 1.6 K, *H* = 0 T; (b) Diagram showing the phase transition from an antiferromagnetic phase to a ferromagnetic phase in applied magnetic field; (c) The two magnetic structures of *M*-TbTaO$_4$ and their evolutions with magnetic fields at 0.50 T, 0.75 T, 1.00 T and 3.00 T. Upper figure shows the antiferromagnetic structure (14.77 magnetic space group), and the lower shows the canted ferromagnetic structure (14.79 magnetic space group). Tb$^{3+}$ ions are shown in blue, tantalum shown in black, and oxygen shown in red. The unit cell is shown in thin grey lines. Magnetic vectors are shown in yellow ('spin down') and blue ('spin up') arrows.

compounds with Ising (easy-axis) or XY (easy-plane) behaviour, the magnetisation often saturates at $g_J \cdot J/2$ or $2g_J \cdot J/3$, respectively, rather than $g_J \cdot J = 9$ $\mu_B$ /f.u. magnetisation. For temperatures slightly above $T_{N1}$, but below 20 K, the magnetisation curve loses the 'S' shape. This suggests thermal energy begins to destabilize the low field antiferromagnetic structure, make it easier for the magnetic moment to align with the field. At high temperatures (above 20 K), the magnetisation response becomes more linear with no distinct features at low fields. This linear response is a typical paramagnetic phase, where magnetic moments align more readily with external field. The differential magnetisation, *dM/dH*, as a function of applied magnetic field is plotted in Fig. 4b at various temperatures. At temperatures below $T_{N1}$, two distinct features could be observed. For data collected at 1.8 K (Fig. 4b inset), A sharp peak appeared at 0.32(2) T, which represents a field-induced spin re-orientation. This peak is likely to a spin-flop transition in the antiferromagnetic phase. A broader peak appeared at 0.76(2) T, which is also observed up to around 4 K. This broader peak suggests a gradual change, likely corresponding to a canted antiferromagnetic phase. As the temperature increases, the sharp low field peak 0.32(2) T is gradually suppressed to 0.09(2) T at 2.25 K and not observed at higher temperatures. This indicates the thermal energy destabilizes the spin flop transition. The broader peak observed at 0.76(2) T remains visible up to 3 K and shows negligible dependence of magnetic field. This suggests it is more robust canted antiferromagnetic phase against thermal fluctuations than low-field sharp peak.

**Specific heat**

The magnetic specific heat of *M*-TbTaO$_4$ at selected fields is shown in Fig. 5a. At the Néel temperature, a sharp peak was observed, this corresponds to the 3D AFM ordering in $d(\chi T)/dT$ plot. On application of a magnetic field, the peak decreases in temperature and becomes broader. Differing from the magnetic susceptibility and isothermal magnetisation measurements, $T_{N2}$ was not seen in the heat capacity data, possibly due to the change in magnetic entropy being too small to detect the insufficient temperature steps resulted in the merge of $T_{N1}$ and $T_{N2}$. The magnetic entropy (**ΔS**) change associated with the transition was obtained by integrating the $C_{mag}/T$ curve from 1.8 to 20.0 K (Fig. 5b). **ΔS** was found to approach 4.0 J mol$^{-1}$ K$^{-1}$ at 0 T and decreased to ~2.4 J mol$^{-1}$ K$^{-1}$ at 2 T. The Ising spins with effective spin of ½ are expected to have the maximum entropy change of $R\ln2$ = 5.76 J mol$^{-1}$ K$^{-1}$ at zero field. The remaining entropy change is assumed to occur below 1.8 K.

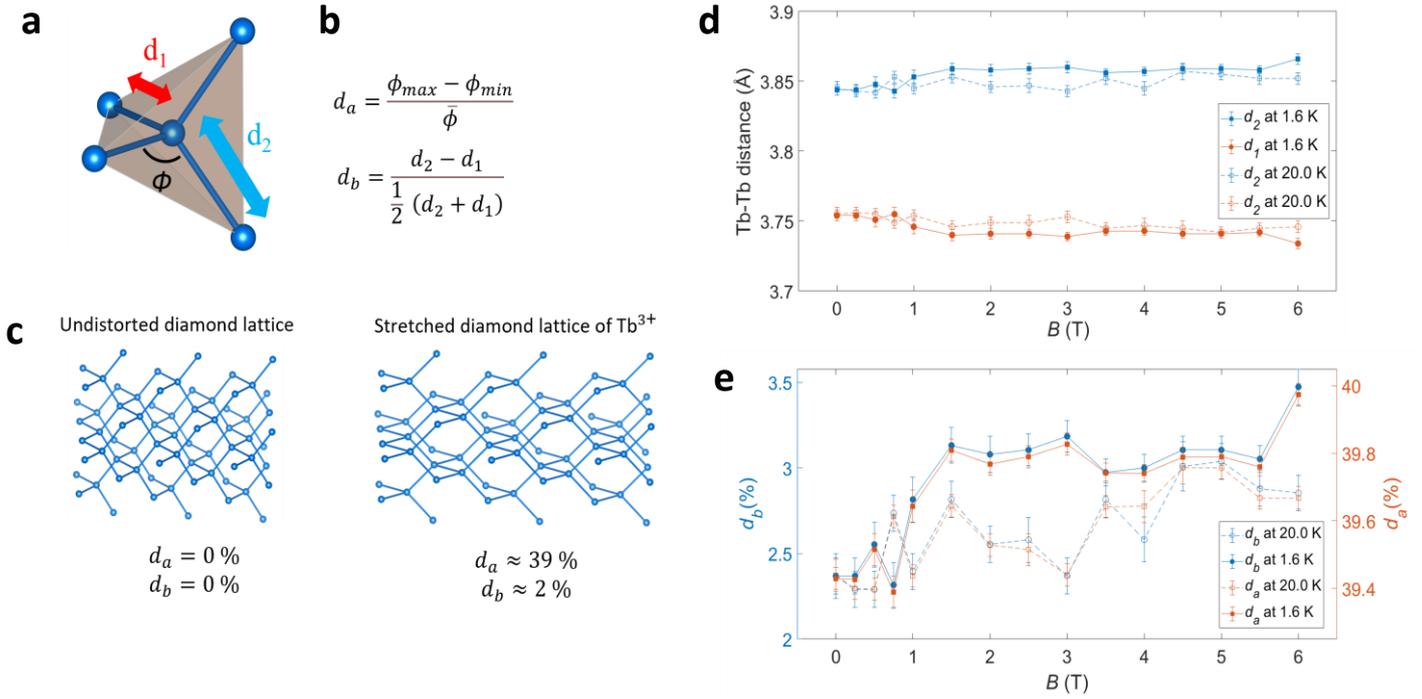

Fig. 8: (a) A single tetrahedral structure of diamond lattice, $d_1$ and $d_2$ are the distance between the nearest-neighbouring atoms and $\phi$ is the angle between bonds; (b) Bond distortion $d_b$ and angle distortion index $d_a$ calculated from $d_1$ and $d_2$ and $\phi$. (c) The undistorted diamond lattice and stretched diamond lattice of $Tb^{3+}$ ions in M-TbTaO$_4$; (d-e) $d_1$, $d_2$, $d_a$ and $d_b$ of $Tb^{3+}$ as a function of magnetic field collected at 20.0 K and 1.6 K from Rietveld refinement of PND collected at $\lambda$ = 2.45 Å.

**Magnetic structure**

High resolution PND data for *M*-TbTaO$_4$ was obtained using the HRPT/SINQ beamline at the PSI. The representative Rietveld refinement of neutron data at zero field is shown in Fig. 6 ab and in-field data are presented in Fig S2 and S3 The results are consistent with the previous literature for TbTaO$_4$[16, 41]. At 20 K, the application of a magnetic field results in the emergence of magnetic Bragg peaks at ~2 T, Fig. 6c-d. The increase in intensity as the field is applied, suggesting a polarization of the $Tb^{3+}$ moments in the paramagnetic phase to align with the applied field. On cooling below the Néel temperature, magnetic Bragg peaks are also observed at 0 T, Fig. 6b, however some of these are suppressed in the presence of an external magnetic field, Fig. 6d. To evaluate these changes more quantitatively, the magnetic peaks were indexed to a commensurate magnetic cell with k=0 in the magnetic space groups P2$_1$'/c (No. 14.77) and P2$_1$'/c'(No. 14.79). Although the crystallographic symmetry is $I2/a$, the magnetic symmetry is reduced to a primitive cell, because the two Tb ions in each unit cell become magnetically inequivalent (one spin up, one spin down). At 1.6 K in zero field the magnetic structure can be described by the P2$_1$'/c (14.77) magnetic space group as reported previously with the Tb spins align perpendicular to the *bc* plane in A-type antiferromagnetic order, as is shown in Fig. 7a. The magnetic moments align colinearly along *a*-axis and form ferromagnetic layers in *ac* plane (Fig. 7b). The ferromagnetic layers are coupled antiferromagnetically along *b*-axis. At zero field, the refined ordered magnetic moment is 7.8(3) $\mu_B$ /Tb$^{3+}$. This is below the theoretical value of 9.72 $\mu_B$ but consistent with 7.5(4) $\mu_B$ which has previously reported by Kelly *et al* [16]. At higher fields of 6 T, the magnetic structure is fully described by the P2$_1$'/c' (14.79) magnetic space groups, with the spins aligned ferromagnetically in the *a* direction, Fig. 7b. At intermediate fields from 0.5 to 3 T, the neutron data can be well fitted with a mixture of two magnetic space groups P2$_1$'/c and P2$_1$'/c', corresponding to two magnetic phases: an antiferromagnetic phase and a ferromagnetic phase, respectively. Note that this approach does not represent a true phase separation but rather is the usual way to describe the spin-canting using TOPAS software. During the refinement, the phase fractions were constrained to remain identical and fixed, while the magnetic moments associated with two magnetic phases were refined independently. For the P2$_1$'/c (14.77) magnetic space group, symmetry restricts the spin to the *a*-axis. The net magnetic moment of the antiferromagnetic phase decreases from 5.4(2) to 0.5(1) $\mu_B$ as magnetic field increases from 0.5 T to 3.0 T, figure S6 In contrast, in the P2$_1$'/c'(No. 14.79) magnetic space group, spin components are allowed along all crystallographic directions. The canted ferromagnetic phase, which emerges at 0.5 T, has a net ferromagnetic component along *a*- and *c*- axes, with an antiferromagnetic component along *b*-axis. Both *b*- and *c*-axis components are progressively suppressed with increasing magnetic field, resulting in a dominant ferromagnetic component along a-axis at 3 T. The net moment of the canted ferromagnetic phase rises from 0.6(2) to 7.1(3) $\mu_B$ along a- axis as the magnetic field ranges from 0.5 to 3.0 T.

These findings imply the formation of a canted antiferromagnetic state from 0.5 to 3.0 T. In this state, the antiferromagnetic alignment of the Tb$^{3+}$ moments begin to cant under the influence of the external magnetic field. At 0 T, the spins are aligned antiferromagnetically along the $a$-axis and are perpendicular to the $bc$ plane. When a magnetic field of 0.5 T is applied, the emergence of net magnetic components along the $b$- and $c$- axes suggests that the spins start to cant towards the $bc$ plane. As the magnetic field increases to 3 T, the $b$ and $c$ components are progressively suppressed, indicating that the spins are reorienting back to align predominantly along the $a$-axis. Eventually, at 6T, the system transitions to ferromagnetic along the $a$-axis.

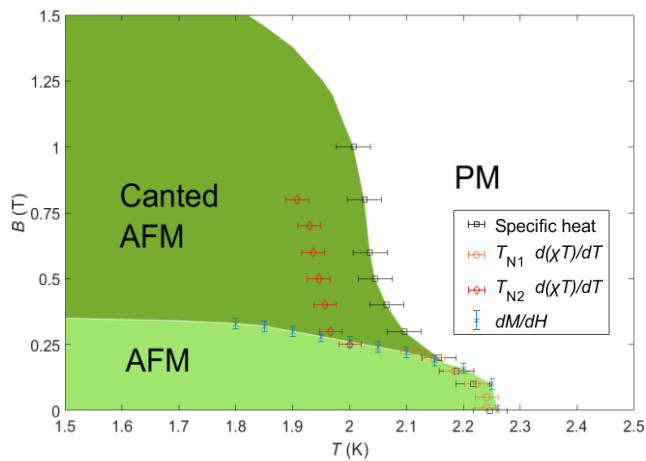

Fig. 9. Proposed magnetic phase diagram for $M$-TbTaO$_4$ as a function of magnetic field and temperature. Transition values derived from the magnetic specific heat, magnetic susceptibility ($T_{N1}$ and $T_{N2}$) and isothermal magnetization are shown as black squares, orange circles, red diamonds, and blue crosses respectively.

In addition to changes in the magnetic structure, we also observed changes in the nuclear structure of $M$-TbTaO$_4$ with magnetic field. In the field range 0-6 T, no significant changes in lattice parameters were observed at 20.0 K. However, at 1.6 K, there is a subtle increase in $a$, $b$, $c$ between 0.75 T and 6 T of ~0.015 %, as is shown in Fig. S6. There are also subtle changes in the refined atomic positions and bond angles on application of a field, notably the decrease of the y coordinates for Tb and O-Tb-O angles with magnetic field at 1.6 K (Figure S7-9).

## Discussion and phase diagram

The magnetic behaviour of $M$-TbTaO$_4$ has been previously discussed with reference to the stretched diamond lattice of Tb$^{3+}$ ions described by the nearest-neighbouring interactions $J_1$ and next-nearest neighbouring interactions $J_2$. It has been reported that the nearest-neighbouring super-exchange in $M$-TbTaO$_4$ predominantly occurs through Tb-O-Tb pathways[16]. These pathways can be divided into two interactions: J$_{1a}$ vector in the $ab$ plane and J$_{1b}$ vector in the $bc$ plane. These interactions depend on the interatomic nearest Tb-Tb distance $d_1$ and $d_2$ respectively (Fig. 8a). At 20 K, there is only small fluctuation in both $d_1$ and $d_2$ in magnetic field. At 1.6 K, $d_1$ exhibits a slight increase with magnetic fields, while $d_2$ shows an opposite tendency (Fig. 8d).

The extent of the stretching or distortion in the diamond-like lattice can be compared using the bond length and the angles between the bonds. In undistorted diamond lattice, all the bond lengths and bond angles are the same (Fig. 8c), and the angles are all equal to 109.5°, while in the monoclinic symmetry there are three different bond lengths and four different angles. Here, we compare the relative deviation from the ideal diamond lattice by using the bond length distortion index $d_a$ and bond angle distortion $d_b$ [16] as is shown in Fig. 8b, where $\phi_{max}$ and $\phi_{min}$ are the largest and smallest angles respectively between Tb-Tb-Tb bonds, respectively. These indices allow us to systematically quantify the structural distortions as the system is subjected to external parameters such as temperature and magnetic field.

Interestingly, the angle and bond distortions exhibit a clear dependency on the applied magnetic field, indicative of an underlying magneto-elastic coupling. At 20 K, only a slight increase in $d_a$ and $d_b$ was observed as the magnetic field increased, but a more pronounced increase in both distortion parameters is observed as the temperature decreased, as is shown in Fig. 8e. At 1.6 K, (T < $T_N$), the bond distortion index $d_b$ increased by approximately 0.6%, and the angle distortion index $d_a$ increased by approximately 1.1% as the magnetic field was varied from 0 to 6 T.

This behaviour can be directly attributed to magneto-elastic coupling, where the interaction between the magnetic moments of the Tb$^{3+}$ ions and the lattice vibrations induce elastic strain in the crystal structure. In this system, the magnetic field alters the interactions and the orientation of magnetic moments, which, in turn, leads to lattice distortions as the crystal attempts to minimize its free energy. Consequently, both bond lengths and bond angles change, reflecting the strong coupling between magnetic order and lattice dynamics. This effect becomes particularly significant at low temperatures below $T_N$ where quantum effects and collective spin ordering dominate the system's behaviour.

The transition temperatures against magnetic fields obtained from $d(\chi T)/dT$, $dM/dH$ and $C_{mag}/T$ have been summarized in the magnetic phase diagram shown in Fig. 9. From 2.25 to 2.11 K (the corresponding field ranges from 0 to 0.225 T), $T_{N1}$ from $d(\chi T)/dT$, $dM/dH$ and $C_{mag}$ overlap with each other, the magnetic specific heat diverges initially below 2.11 K, and ~1.967 K (the corresponding field =0.25 T) for $T_{N2}$. The divergency below Néel temperature separates the phase diagram into three regimes, which we classify as AFM, canted AFM and FM phases from our PND study. AFM phase refers to the phase that the Tb spins align perpendicular to $bc$ plane, whilst in the canted AFM phase, the Tb spins are gradually canted due to the increasing field but remain the AFM ordering, finally entering the FM phase at higher fields. Further study of the hard and easy axis will require anisotropic magnetisation measurements on single crystals.

## Conclusions

We have systematically investigated the magnetic properties of polycrystalline $M$-TbTaO$_4$ as a function of temperature and applied magnetic field. Three-dimensional antiferromagnetic ordering is observed at $T_{N1}$ and consistent with the previous literatures. We discover, for the first time, a second Néel temperature $T_{N2}$ in $d(\chi T)/dT$. We find that both Néel temperatures ($T_{N1}$ and $T_{N2}$) in $M$-TbTaO$_4$ could be suppressed on application of a magnetic field up to 0.8T. Overall, this suggests a detailed three-regime phase diagram in $M$-TbTaO$_4$ below its transition temperature.

We have also studied the evolution of magnetic structure by powder neutron diffraction measurements below and above $T_N$, under varying magnetic fields. Below the AFM transition temperature of 2.25 K, we observed the appearance of magnetic peaks which were suppressed by magnetic field. We find that the magnetic moments of Tb$^{3+}$ ions, which are initially aligned parallel to the $a$-axis in Néel AFM order, cant towards the $bc$ plane on application of the external magnetic field and eventually form a FM order parallel to the $a$-axis again

The Tb$^{3+}$ sites in in $M$-TbTaO$_4$ have been reported to from an elongated and stretched diamond lattice. Notably, we have observed a slight increase in both angle and bond distortion of this stretched diamond lattice from 0 to 6 T at 1.6 K. This indicates that the antiferromagnetic ordering in $M$-TbTaO$_4$ potentially be linked to a magneto-elastic coupling effect.

Future work including dielectric measurements would be highly required to unveil the potential coupling effect between magnetic order and nuclear order. The magnetic measurement on single crystal would also be essential to quantify the possible magnetic anisotropy along different axes in $M$-TbTaO$_4$.

## Author contributions

X. Zhang: conceptualization, data curation, formal analysis, investigation, validation, visualization, writing original draft, writing– review & editing. N. Kelly: conceptualization, data curation, formal analysis, investigation, writing– review & editing. D. Sheptyakov: data curation, investigation, resources. C. Liu: data curation, investigation, resources, writing– review & editing. S. Deng: data curation, investigation. S. Saxena: conceptualization, formal analysis, investigation, supervision, project administration, funding acquisition, resources, writing review & editing. S. Dutton: conceptualization, formal analysis, investigation, supervision, project administration, funding acquisition, resources, writing review & editing.

## Data availability

The datasets that support the findings of this study, including those from magnetic measurement, heat capacity measurement, Powder X-ray and Neutron diffraction, are available at https://doi.org/10.17863/CAM.113572.

## Conflicts of interest

There are no conflicts to declare.

## Acknowledgements

This work was financially supported by the funding from the Department of Business, Energy, and Industrial Strategy (BEIS) (Grants No. G115693.). This work is partly based on experiments performed at the Swiss spallation neutron source SINQ, Paul Scherrer Institut, Villigen, Switzerland. The authors gratefully acknowledge the technical support provided at Paul Scherrer Institut.

SUPPLEMENTARY INFORMATION

for

Magnetoelastic coupling in the stretched diamond lattice of TbTaO$_4$

Xiaotian Zhang,*[a] Nicola D. Kelly,[ab] Denis Sheptyakov,[c] Cheng Liu,[a] Shiyu Deng,[ad]

Siddharth S. Saxena*[a] and Siân E. Dutton *[a]

[a] Cavendish Laboratory, University of Cambridge, JJ Thomson Avenue, Cambridge, CB3 0HE, United Kingdom.

[b] Department of Chemistry, University of Oxford, Mansfield Rd, Oxford OX1 3TA, United Kingdom.

[c] Laboratory for Neutron Scattering and Imaging, Paul Scherrer Institut (PSI), Forschungsstrasse 111, 5232 Villigen PSI, Switzerland.

[d] Institut Laue-Langevin (ILL), 71 Avenue des Martyrs, 38000 Grenoble, France.

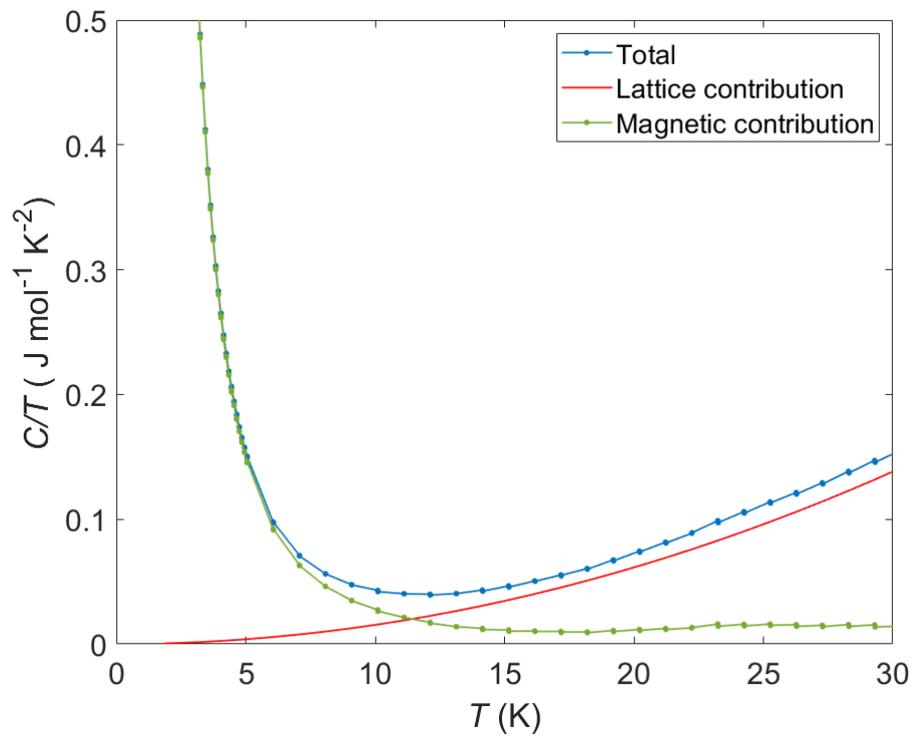

Fig. S1 Heat capacity data for *M*-TbTaO$_4$. Blue dots: total heat capacity (after subtraction of the Ag contribution from the raw data). Red line: Estimated lattice contribution using the Debye model with $\Theta_d$=370 K[1]. Green dots: magnetic contribution, $C_{mag}=C_{total}-C_{lattice}$.

Table S1: Refined Lattice parameters and atomic positions of *M*-TbTaO$_4$ from powder X-ray diffraction at room temperature. The O atomic positions were fixed at value from previous neutron diffraction data[2, 3]. Both Tb$^{3+}$ and Ta$^{5+}$ ions lie on (0.25, y , 0) sites.

| Unit cell | |
|---|---|
| Space group | *I2/a* |
| *a*/ Å | 5.38264(15) |
| *b*/ Å | 11.0182(3) |
| *c*/ Å | 5.06692(13) |
| *β*/° | 95.6844(14) |
| Volume/ Å$^3$ | 299.025(14) |
| *R$_{wp}$*/% | 7.21 |
| *χ2* | 2.10 |
| Atomic positions | |
| *yTb* | 0.6180 (3) |
| *yTa* | 0.1502 (2) |
| *xO1* | 0.090 |
| *yO1* | 0.461 |
| *zO1* | 0.252 |
| *xO2* | -0.003 |
| *yO2* | 0.717 |
| *zO2* | 0.290 |

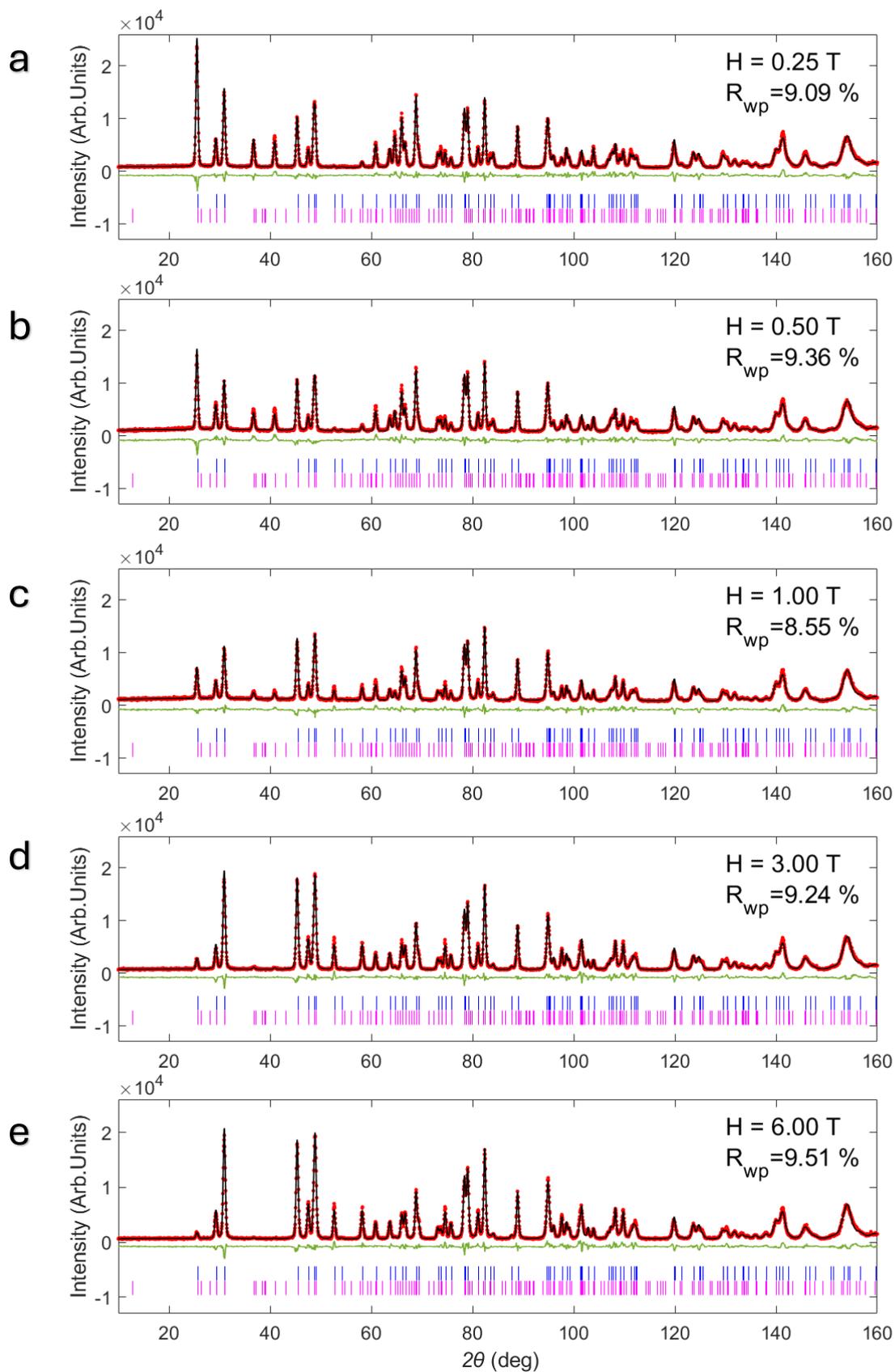

Fig. S2 (a-e) Refined PND data of *M*-TbTaO4 collected at 1.6 K with 0.25 T, 0.50 T, 1 T, 3 T and 6 T. Red dots, experimental data; black line, calculated intensities; green line, difference pattern; tick marks, nuclear (blue), magnetic (pink) Bragg reflection positions. λ = 2.45 Å.

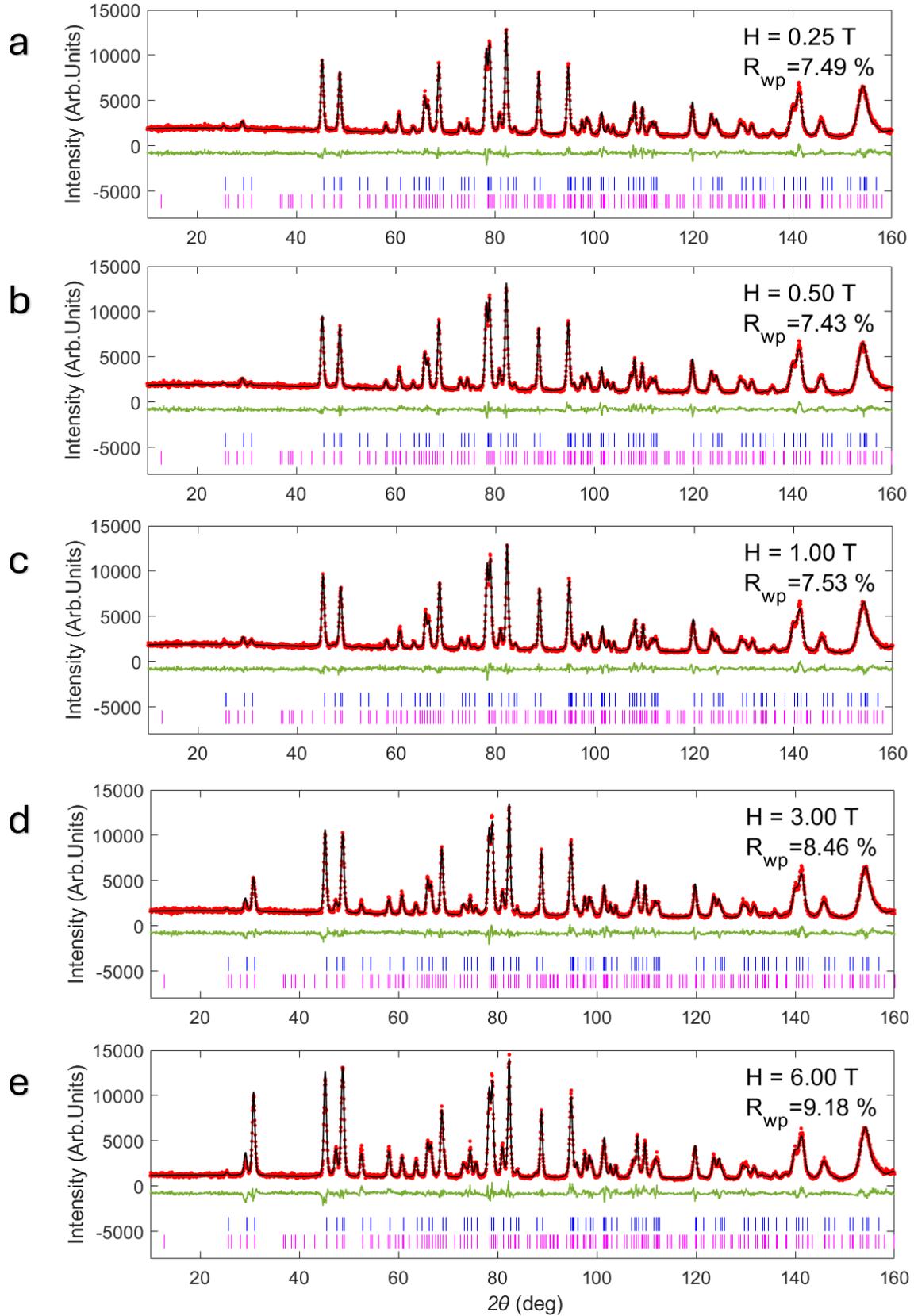

Fig. S3 (a-e) Refined PND data of *M*-TbTaO$_4$ collected at 20.0 K with 0.25 T, 0.50 T, 1 T, 3 T and 6 T. Red dots, experimental data; black line, calculated intensities; green line, difference pattern; tick marks, nuclear (blue), magnetic (pink) Bragg reflection positions. λ = 2.45 Å.

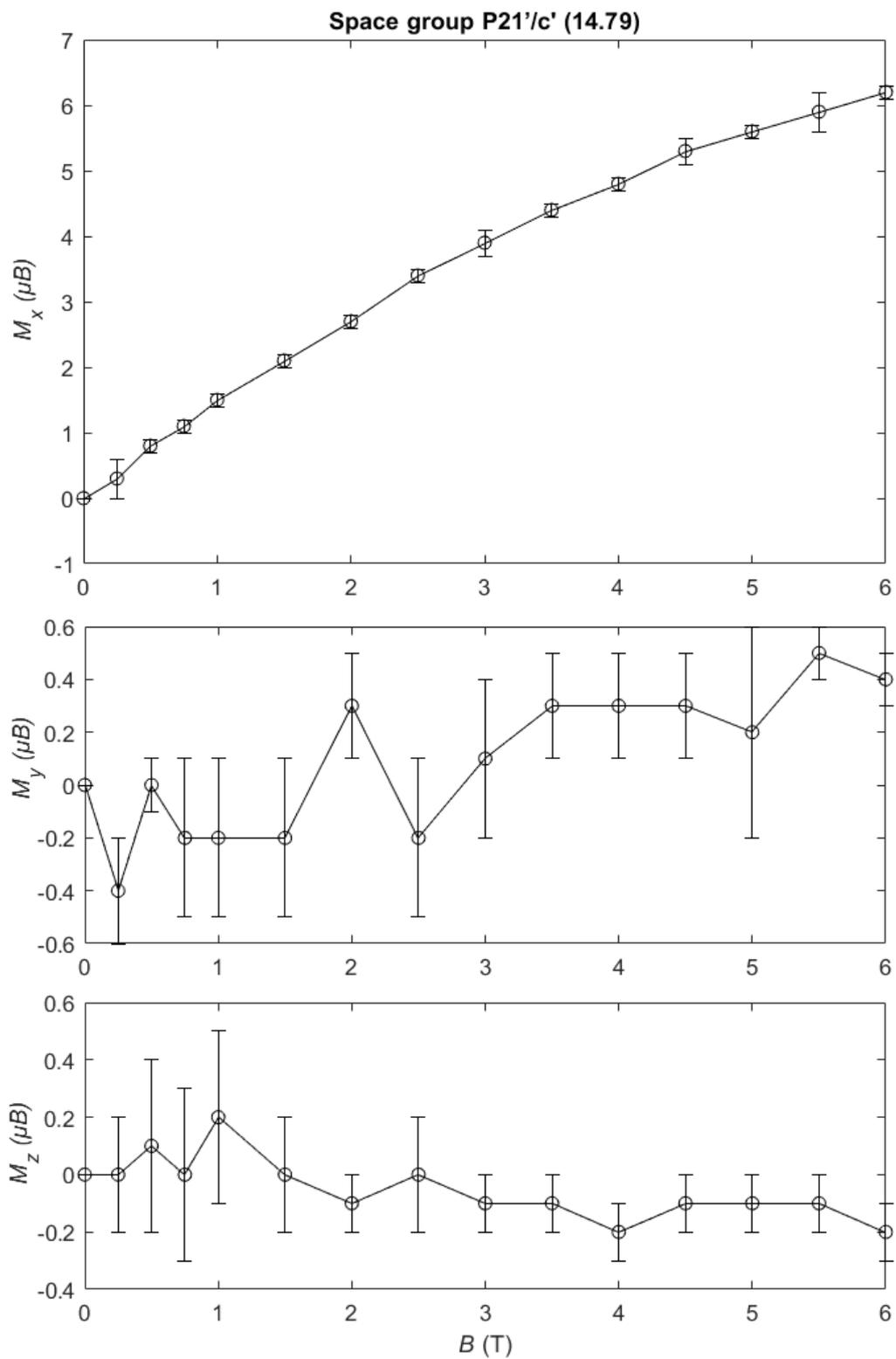

Fig. S4 Evolution of the $Tb^{3+}$ lattice-coordinate magnetic moments along x, y and z axes as a function of magnetic field at 20.0 K from powder neutron diffraction (PND) data.

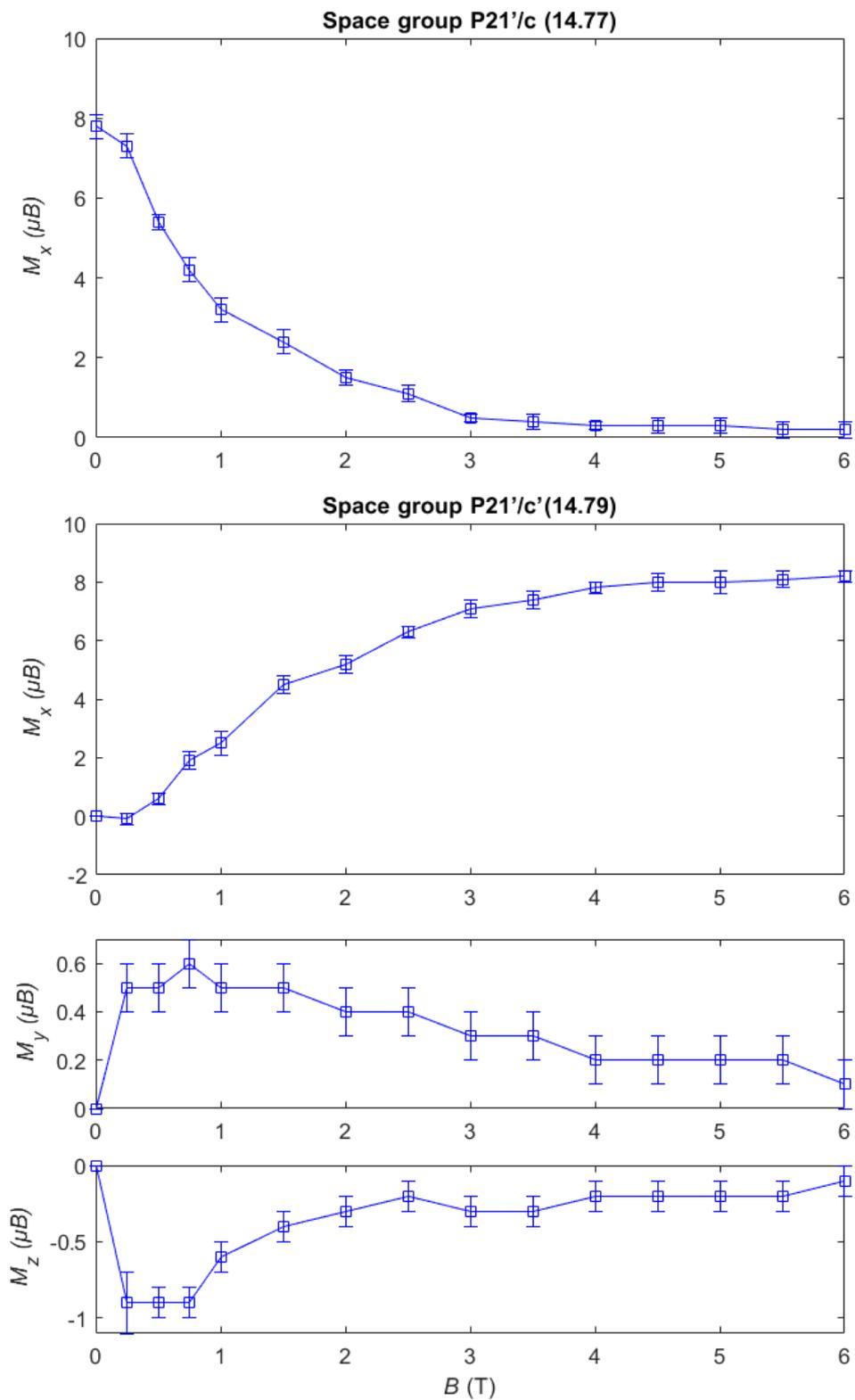

Fig. S5 Evolution of the $Tb^{3+}$ lattice-coordinate magnetic moments along x, y and z axes as a function of magnetic field at 1.6 K from powder neutron diffraction (PND) data. The magnetic structure is modelled with a mixing of two magnetic space group. The phase fractions are fixed and identical to each other, while the magnetic moments are refined independently.

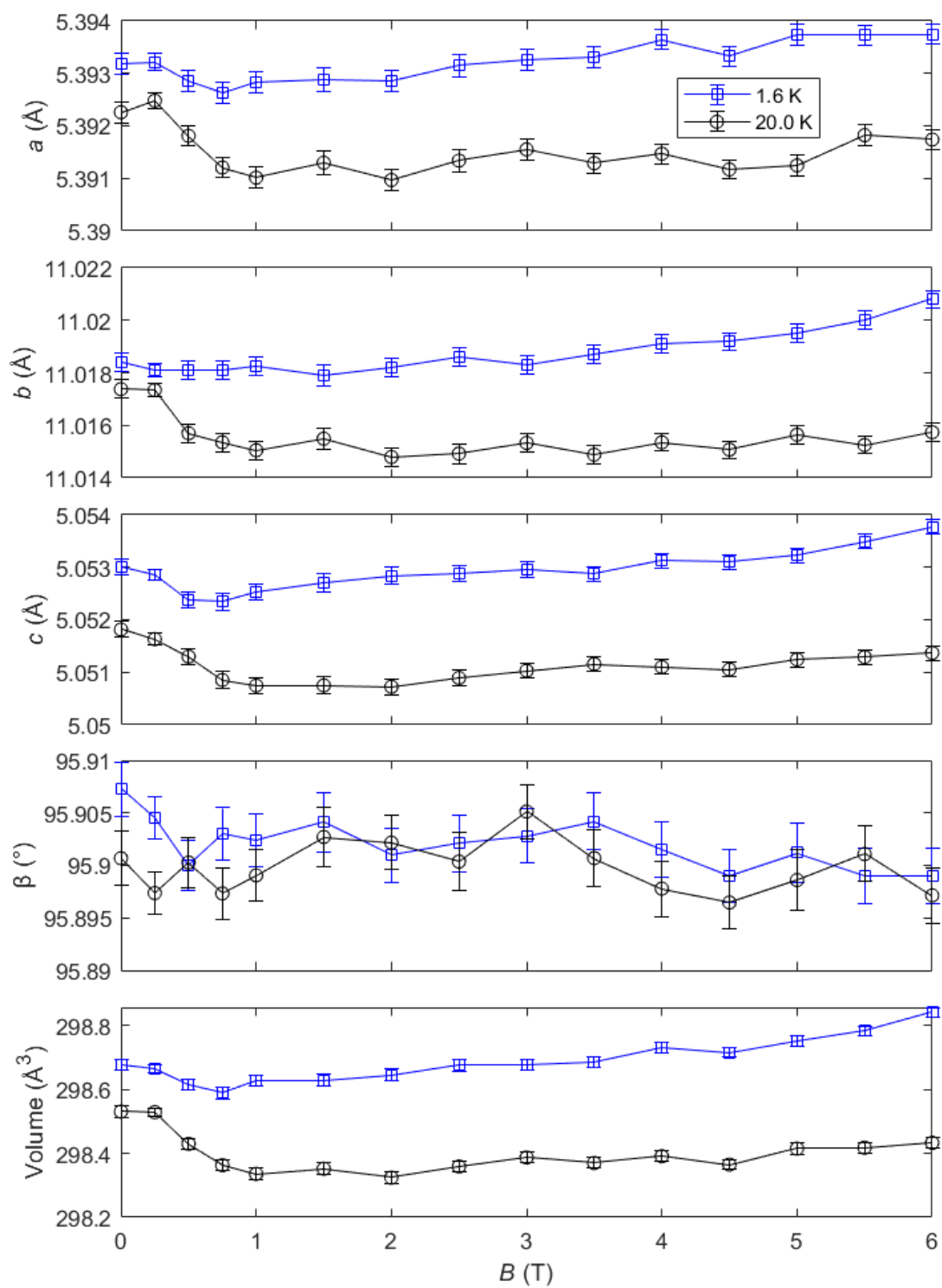

Fig. S6 Refined lattice parameters of *M*-TbTaO$_4$ at 1.6 K and 20K as a function of magnetic field. Obtained from Rietveld refinement of powder neutron diffraction (PND) data collected on HRPT, PSI. λ = 2.45 Å.

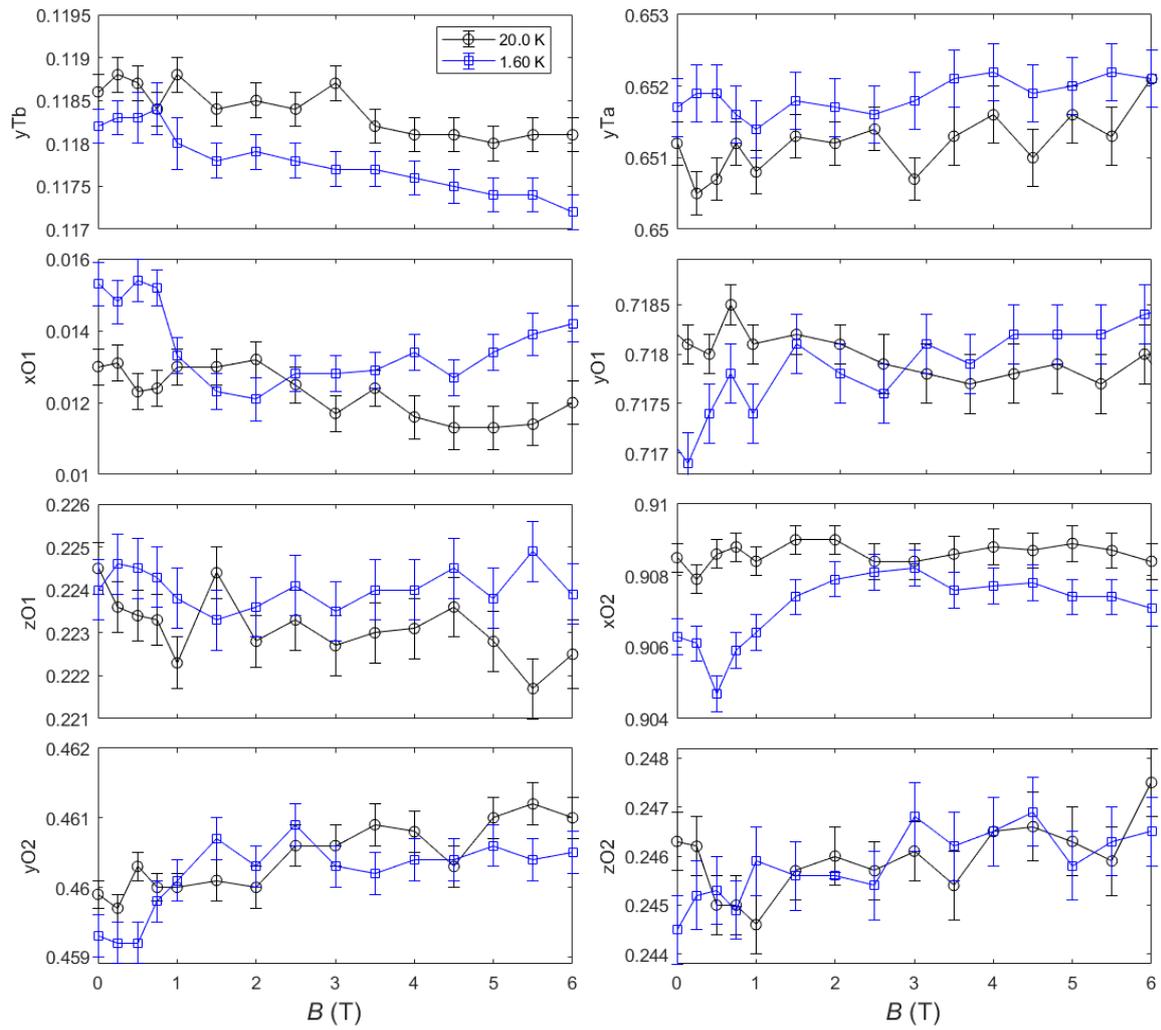

Fig. S7 Refined atomic positions of *M*-TbTaO$_4$ at 1.6 K and 20K as a function of magnetic field. Obtained from Rietveld refinement of powder neutron diffraction (PND) data collected on HRPT, PSI. λ = 2.45 Å.

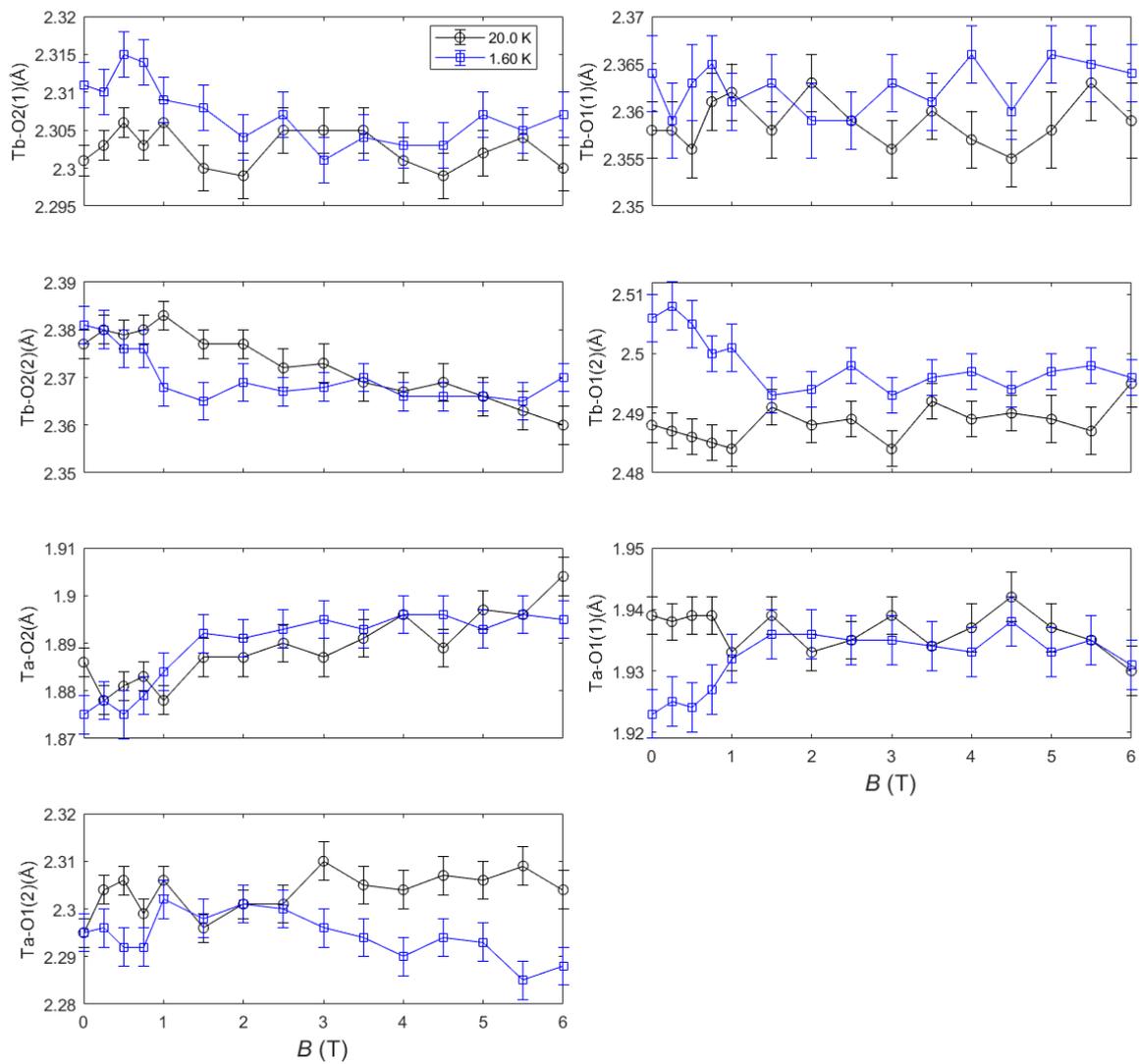

Fig. S8 Refined interatomic distances of *M*-TbTaO$_4$ at 1.6 K and 20K as a function of magnetic field. Obtained from Rietveld refinement of powder neutron diffraction (PND) data collected on HRPT, PSI. λ = 2.45 Å.

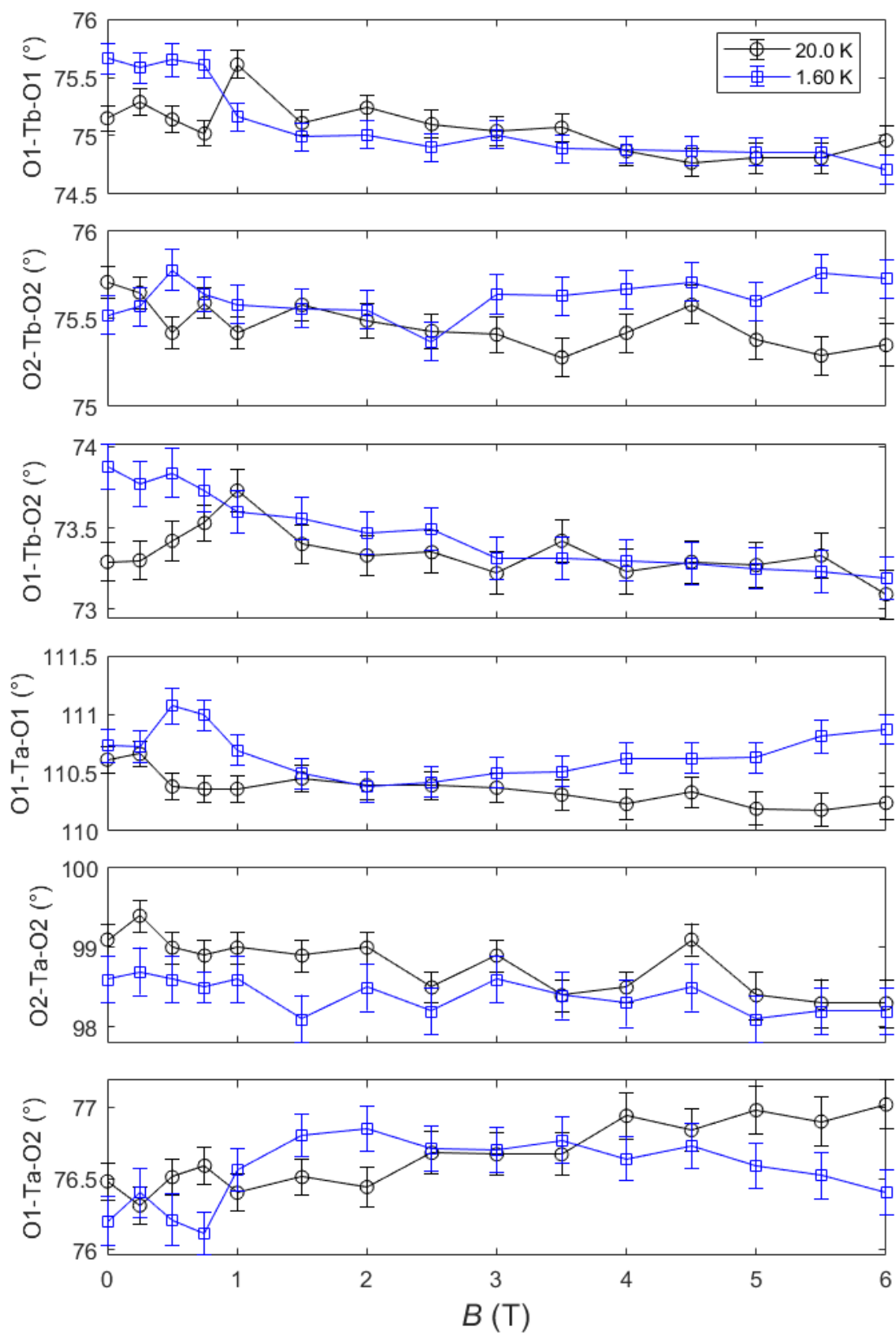

Fig. S9 Refined bond angles of *M*-TbTaO$_4$ at 1.6 K and 20K as a function of magnetic field. Obtained from Rietveld refinement of powder neutron diffraction (PND) data collected on HRPT, PSI. λ = 2.45 Å.